# An Empirical Analysis of Success Factors in the Adoption of the Scaled Agile Framework – First Outcomes from an Empirical Study

Dilshat Salikhov, Giancarlo Succi, Alexander Tormasov

Innopolis University

Innopolis, Russia

d.salikhov@innopolis.ru

**Abstract**

Agile methodologies are used for improving productivity and quality of development originally created for small teams. However, now they are expanding to larger organizations, for which "scaled up" approaches have been proposed. This study presents the preliminary outcomes from a survey on the effects of the Scaled Agile Framework (SAFe), which is considered the most used of such approaches. The apparent advantages and limitations are discussed along with the lines for future research.

## 1 Introduction

Agile methods started emerging in mid-1990s as an alternative to traditional, "plan-driven" methods [1–4]. It was a response to a crisis of the plan methods that do not provide enough flexibility in the modern business environment [5]. It was initially developed for small and collocated teams of no more than 20 people [5]. From the very beginning, agile practitioners reported a problem about synchronization in agile teams with the rest of the organization, that usually works using plan-driven approaches. This is a common problem throughout organizations when moving to a new approaches [6–9]. However, it becomes particularly acute for agile, for the intrinsically different approach in structuring the whole operations. Consequently, large organizations started adopting agile practices and structures at an organizational level [10].

The scaling of agile methods brings multiple challenges: synchronization and coordination and communication and interfaces between different teams [11]. Several scaling frameworks have emerged during the last few years: SAFe, LeSS, DAD[10]. The most popular one, according to the agile annual report, is SAFe [12].

SAFe is a scalable and configurable framework that helps organizations deliver solutions. It describes the roles, responsibilities, structure and activities using a combination of agile, lean development, DevOps and system thinking[13].

It has a hierarchical structure and consists of four levels: team, program, large solution and portfolio. The team level works like the usual Scrum team, however, teams are allowed to freely choose between Scrum, Kanban or Scrumban. The program is a collection of levels. The large solutions coordinate multiple programs. The portfolio-level management dictates the direction and priorities of the teams [13].

However, there are alleged limitations of SAFe. SAFe might be too complicated a framework and hard to understand and the framework creates top-down organizational control that leads to a strict organization structure with lesser flexibility [14].

This paper aims at summarizing the early findings of an investigation over the factors that affect the success of SAFe adoption. An additional aim is to understand if SAFe is still agile, how it impacts the agility in the company and what benefits and drawbacks it brings to the work. The key research questions that we are targeting are:

> **RQ1:** What are the main factors of SAFe's success or failure in large organizations?
> **RQ2:** What are the causes that may make SAFe not agile?
> **RQ3:** What are the benefits and drawbacks of SAFe?

This paper is organized as follows. Section 2 briefly summarises the state of the art. Section 3 outlines our research instrument. Section 4 presents the results we have obtained and Section 5 discusses them. Section 6 presents some conclusions and outlines our future research.

## 2 State of the art

The Agile Annual Report showed that the most used scaling framework is SAFe – 30% of respondents use it for scaling agile [15]. Several studies have been conducted on scaling agile frameworks, including case studies, literature reviews and surveys. The issue of an alleged non-agility of SAFe, despite being mentioned in gray literature, has not been studied in research papers so far as we could not find peer-reviewed studies on this topic, apart from some references in lateral works [16, 17] and an explicit mention in the Large-Scale Agile workshop [14]. The current state is presented by grey literature, hence, the current results are based on practitioners, judgement and opinions. Jeffries [18] and Evans [19] presented their blogposts where they shared their opinions on why SAFe violates agile values.

Furthermore, there are a few peer-reviewed researches about the consequences of SAFe application. Laanti and Kettunen conducted a survey about SAFe adoption in Finnish organizations. One hundred eleven results were collected and reported a huge variety of benefits and drawbacks. The limitation of this survey is the population – only 5 percent of the participants were pure SAFe users [20]. Additionally, there are few case studies, not rigorous enough and hard to generalize [21].

## 3 Research instruments

The survey based on interviews was chosen as the methodology for this study because the research questions are in the form of

"what" and also the fact that our research nature is exploratory [21]. We also chose interview as an approach for conducting survey.

The interviews were standardized, open-ended, prototyped with two rounds of pilots and then finalized [22, 23]. The structure of the interview is organized in three parts: the first consists of background questions for retrieving some contextual information about participants, the second consists of Likert scale questions on a scale from 1 to 5 on why SAFe breaks agile values based on the outcome of the SLR. The third part is presented through open-ended questions. The numeric answer of the second part were analised via median and standard deviation [24]. The open-ended questions of the third part were coded according to Creswell [23]: windowing the data, organizing it, coding it, definition of themes, organization of themes, and extraction of the lesson learnt.

Triangulation by data source was used for improving the quality of the results. We tried to access participants from different companies in different positions, from team to top management level. [25]. We selected participants, experienced SAFe developers, via LinkedIn, using SAFe communities, company search (that we already know of as adopting or have adopted SAFe) and searching for specific SAFe roles. Moreover, we have followed the rigorous practices defined by scientific community throughout the years [26–46]. To analyse the resulting data we have used open source tools [47–50]

## 4 Results

We have got 21 responses out of 136 requests. Employees of sixteen organizations participated in this research. We approached SAFe practitioners from all levels but the respondents were 7 for team, 13 for program, 2 for large solution and none for portfolio. We had minimal coverage error because the requested practitioners were from all levels. However, the nonresponse error was quite high as nobody from the portfolio level answered.

We identified eleven success factors that support the adoption of SAFe (Figure 1). All the top five success factors, but competent staff, has complementing items in interfering factors list, showing that support their importance. Further, our interfering factors (Figure 2) are clearly overlapped with those from the 13th Agile State Report [15]. Moreover, we identified the factors that were previously was not present in the literature: appropriate culture, moving responsibilities down and communication between layers. Moving responsibilities down was explained by one of the practitioners as "pushing down decision-making authority to the lowest responsible levels to empower team members and create a stronger sense of ownership."

The analysis of results is shown in Table 1. The results show that participants agree that SAFe is conformant to agile but not strongly. Six out of twenty-one respondents explained their disagreement with first statement due to the fact that SAFe is just a set of practices that are good to follow. One of the respondents added that SAFe allows all levels of an organization to choose which path to focus on to achieve their goals. The disagreements with the second statement are similar to the previous. However, there is an additional point from one of the participants: in the spirit of Shu-Ha-Ri [51], it best to first adhere to the guidance strictly and then learn how to adapt it. For the third statement, five participants referred to the Inspect

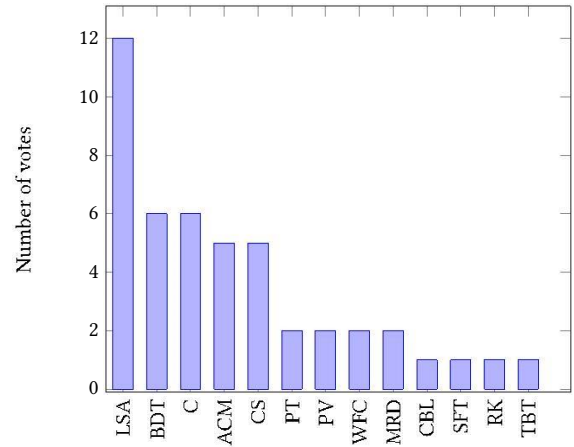

**Figure 1: Success factors**
LAS: leadership support in adoption, BDT: business and development together, C: coaching, ACM: appropriate culture and mindset, CS: competent staff, PT: pilot teams, PV: product vision, WFC: well-founded choice, MRD: moving responsibilities down, CBL: communication between layers, SFT; support from teammates, RK: relevant knowledge, TBT: trust between teams.

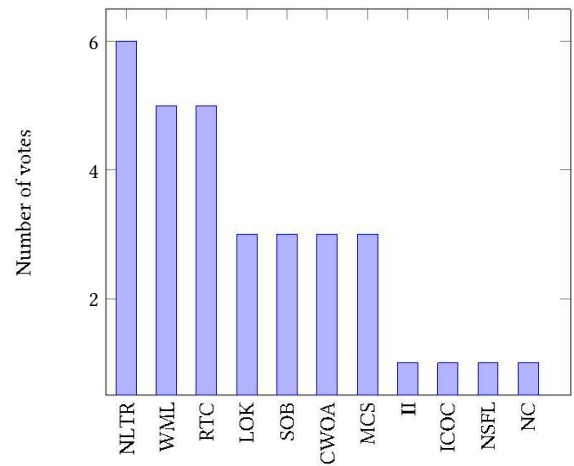

**Figure 2: Interfering factors**
NLTR: need of a lot of time or resources, WML: wrong mindset of leaders, RTC: resistance to change, LOK: lack of knowledge, SOB: scepticism of business, CWOA: conflict with old approach, MCS: more complex structure, II: incorrect implementation, ICOC: inappropriate culture of company, NSFL: no support from leaders, NC: no coaching.

& Adapt process of SAFe. There is no consensus around the last statements.

We identified several benefits and drawbacks (Figures 3 and 4 respectively). In terms of benefits, we have what was expected in terms of better productivity, better handling of dependencies, more coordination, and sharing a common vision. Overall, we notice that the ones present in the SAFe white paper [13] are also mentioned in the survey apart from faster time to market. The most commonly identified drawbacks are the requirement of more resources, the complexity, and the lack of autonomy. Also some people mentioned

|   | Median | Std. |
|---|---|---|
| SAFe provides a comprehensive model of management that complicates the whole process | 2 | 1.08 |
| SAFe focuses more on mandating specific methods rather than fitting them to the context | 2 | 0.93 |
| SAFe does not support teams in learning and improving the process based on experience. | 2 | 1.06 |
| Alignment in SAFe means "everyone does what is defined by management," leading to the inability of teams to optimize their operations. | 2 | 0.94 |

Table 1: Agreements with statements on SAFe on a Lippert Scale from 1 ("strongly disagree") to 5 ("strongly disagree"); 2 means "disagree."

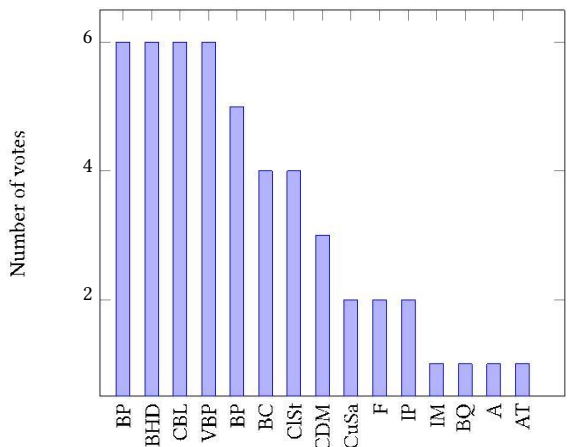

Figure 3: Benefits

**BP:** better productivity, **BHD:** better handling dependencies, **CBL:** coordination between levels, **VBP:** vision of the big picture, **BP:** better predictability, **BC:** better communication, **ClSt:** clear structure, **CDM:** centralized decision making, **CuSa:** customer satisfaction, **F:** flexibility, **IP:** improved prioritization, **IM:** increased motivation, **BQ:** better quality, **A:** alignment, **AT:** autonomous teams.

decreased productivity making the point of productivity controversial, and a small but not insignificant number of people mentioned more stress and customers dissatisfaction.

## 5 Discussion

The purpose of this research was to study the consequences of agile adoption, understand what can help and interfere during adoption and explore non-agility nature of SAFe.

We identified the importance of leadership in adoption process. The factors related to leadership are most important success factor(*Leadership support and adoption*) and second popular among interfering (*Wrong mindset of leaders*) factors. The leadership should not only support adoption but also adapt themselves to new work methods.

Furthermore, there was one common idea among participants about non-agility of SAFe. The idea is that SAFe becomes non-agile in case of incorrect implementation. It is similar to team level agile methods such as Scrum, because Scrum also can be implemented wrongly. Some of the practitioners calls such implementation ScrumFall [52].

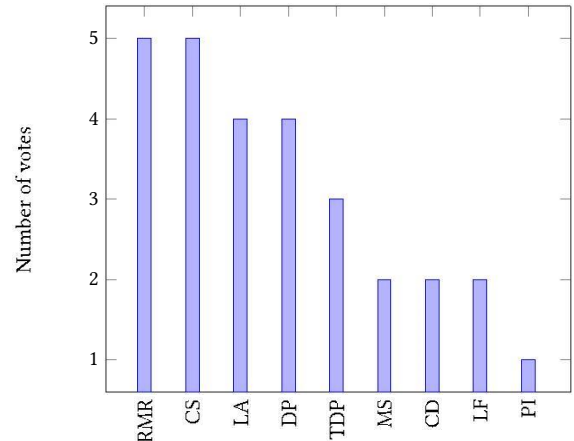

Figure 4: Drawbacks

**RMR:** requires more resources, **CS:** complex structure, **LA:** lack of autonomy, **DP:** decreased productivity, **TDP:** top-down planning, **MS:** more stress, **CD:** customer dissatisfaction, **LF:** lack of flexibility, **PI:** prioritization of issues.

This research can serve as a base for future quantitative researches about how to adopt SAFe and what specific problems may arise with this framework. Furthermore, it can help studies that compare SAFe with different scaling frameworks such as LeSS or RAGE.

## 6 Conclusion

In this qualitative study, we researched what factors helped to adopt the Scaled Agile framework and what challenges organizations and their employees face during the adoption process. We have collected information through a set of structured interviews in different companies.

Our results support the existing knowledge in this area that consists of several case studies. We found that *leadership support and adoption* are the most important factors in SAFe transformation. Additionally, it was supported by one of the most common challenges of *wrong mindset of leaders*. We also tested the hypotheses of agile practitioners about non- agility of SAFe. The results suggest that most of them failed. We also identified several benefits and drawbacks of SAFe adoption. Moreover, we analyzed how these consequences were measured.

It would be now interesting to expand the research with a deeper understanding of the answers and considering specific application domains, such as mobile [53, 54].

## References

[1] J. Kivi, D. Haydon, J. Hayes, R. Schneider, and G. Succi, "Extreme programming: a university team design experience," in *2000 Canadian Conference on Electrical and Computer Engineering. Conference Proceedings. Navigating to a New Era (Cat. No.00TH8492)*, vol. 2, May 2000, pp. 816–820 vol.2.

[2] A. Sillitti, T. Vernazza, and G. Succi, "Service Oriented Programming: A New Paradigm of Software Reuse," in *Proceedings of the 7th International Conference on Software Reuse*. Springer Berlin Heidelberg, April 2002, pp. 269–280.

[3] A. Janes and G. Succi, *Lean Software Development in Action*. Heidelberg, Germany: Springer, 2014.

[4] S. Abdalhamid and A. Mishra, "Adopting of agile methods in software development organizations: Systematic mapping," *TEM Journal*, vol. 6, no. 4, pp. 817–825, 2017. [Online]. Available: https://dx.doi.org/10.18421/TEM64-22

[5] S. K. Weiss and P. Brune, "Crossing the Boundaries – Agile Methods in Large-Scale, Plan-Driven Organizations: A Case Study from the Financial Services Industry,"

2017, pp. 380–393. [Online]. Available: http://link.springer.com/10.1007/978-3-319-59536-8{_}24
[6] A. Valerio, G. Succi, and M. Fenaroli, "Domain analysis and framework-based software development," *SIGAPP Appl. Comput. Rev.*, vol. 5, no. 2, pp. 4–15, Sep. 1997.
[7] J. Clark, C. Clarke, S. De Panfilis, G. Granatella, P. Predonzani, A. Sillitti, G. Succi, and T. Vernazza, "Selecting components in large cots repositories," *Journal of Systems and Software*, vol. 73, no. 2, pp. 323–331, 2004.
[8] W. Pedrycz and G. Succi, "Genetic granular classifiers in modeling software quality," *Journal of Systems and Software*, vol. 76, no. 3, pp. 277–285, 2005.
[9] W. Pedrycz, B. Russo, and G. Succi, "Knowledge Transfer in System Modeling and Its Realization Through an Optimal Allocation of Information Granularity," *Appl. Soft Comput.*, vol. 12, no. 8, pp. 1985–1995, Aug. 2012.
[10] D. C. D. Gerster, ""Agile Meets Non-Agile": Implications of Adopting Agile Practices at Enterprises," *Conference: 24th American Conference on Infomation Systems*, 2018. [Online]. Available: https://www.researchgate.net/publication/331963482{_}Agile{_}Meets{_}Non-Agile{_}Implications{_}of{_}Adopting{_}Agile{_}Practices{_}at{_}Enterprises
[11] M. Shameem, C. Kumar, B. Chandra, and A. A. Khan, "Systematic review of success factors for scaling agile methods in global software development environment: A client-vendor perspective," in *Proceedings - 2017 24th Asia-Pacific Software Engineering Conference Workshops, APSECW 2017*, vol. 2018-Janua, 2018, pp. 17–24. [Online]. Available: http://dl.acm.org
[12] VersionOne, "Annual state of agile," 2019. [Online]. Available: https://www.stateofagile.com/#ufh-i-521251909-13th-annual-state-of-agile-report/473508
[13] Scaled Agile, "Scaled agile framework - safe for lean enterprises," *SAFe for enterprises*, no. November, 2018. [Online]. Available: https://www.scaledagile.com/resources/safe-whitepaper/https://www.scaledagile.com/resources/safe-whitepaper/{%}0Ahttps://www.scaledagileframework.com/{%}0Ahttp://www.scaledagileframework.com/
[14] T. Dingsøyr, N. B. Moe, and H. H. Ohlsson, "Towards an understanding of scaling frameworks and business agility," in *Proceedings of the 19th International Conference on Agile Software Development Companion - XP '18*. ACM Press, 2018. [Online]. Available: http://dl.acm.org/10.1145/3234152.3234176
[15] VersionOne, "13th annual state of agile report," 2018. [Online]. Available: https://www.stateofagile.com/?{_}ga=2.142827005.93339774.1571472163-1493396725.1571472163{#}ufh-i-521251909-13th-annual-state-of-agile-report/473508
[16] M. Paasivaara, "Adopting SAFe to Scale Agile in a Globally Distributed Organization," in *2017 IEEE 12th International Conference on Global Software Engineering (ICGSE)*. IEEE, may 2017, pp. 36–40. [Online]. Available: http://ieeexplore.ieee.org/document/7976685/
[17] J. Pries-Heje and M. M. Krohn, "The SAFe way to the agile organization," in *ACM International Conference Proceeding Series*, ser. XP '17, vol. Part F1299. New York, NY, USA: ACM, 2017, pp. 18:1–18:3. [Online]. Available: http://doi.acm.org/10.1145/3120459.3120478
[18] Ron Jeffries, "SAFe - Good But Not Good Enough," 2014. [Online]. Available: https://ronjeffries.com/xprog/articles/safe-good-but-not-good-enough/
[19] Kyle Evans, "The Major Problems with SAFe," 2019. [Online]. Available: https://productcoalition.com/the-major-problems-with-safe-1e797f7e48f8
[20] M. Laanti, P. K. I. C. o. A. Software, and U. 2019, "SAFe adoptions in Finland: a survey research," *Springer*. [Online]. Available: https://link.springer.com/chapter/10.1007/978-3-030-30126-2{_}10
[21] R. K. Yin, "Applications of case study research," *Applied Social Research Methods Series*, 2013.
[22] N. Bradburn, S. Sudman, and B. Wansink, "Asking Questions: The Definitive Guide to Questionnaire Design," *The Definitive Guide to Questionnaire Design - For Market Research, Political Polls, and Social and Health Questionnaires*, 2004.
[23] J. W. Creswell, *Qualitative Inquiry and Research Design*, 2013.
[24] H. N. Boone and D. A. Boone, "Analyzing Likert data," *Journal of Extension*, 2012.
[25] M. Patton, "Qualitative research and evaluation methods," *http://lst-iiep.iiep-unesco.org/cgi-bin/wwwi32.exe/[in=epidoc1.in]/?t2000=018602/(100)*, vol. 3, 01 2002.
[26] G. Marino and G. Succi, "Data Structures for Parallel Execution of Functional Languages," in *Proceedings of the Parallel Architectures and Languages Europe, Volume II: Parallel Languages*, ser. PARLE '89. Springer-Verlag, June 1989, pp. 346–356.
[27] F. Maurer, G. Succi, H. Holz, B. Kötting, S. Goldmann, and B. Dellen, "Software Process Support over the Internet," in *Proceedings of the 21st International Conference on Software Engineering*, ser. ICSE '99. ACM, May 1999, pp. 642–645.
[28] T. Vernazza, G. Granatella, G. Succi, L. Benedicenti, and M. Mintchev, "Defining Metrics for Software Components," in *Proceedings of the World Multiconference on Systemics, Cybernetics and Informatics*, vol. XI, July 2000, pp. 16–23.
[29] G. Succi, L. Benedicenti, and T. Vernazza, "Analysis of the effects of software reuse on customer satisfaction in an RPG environment," *IEEE Transactions on Software Engineering*, vol. 27, no. 5, pp. 473–479, 2001.
[30] G. Succi, J. Paulson, and A. Eberlein, "Preliminary results from an empirical study on the growth of open source and commercial software products," in *EDSER-3 Workshop*, 2001, pp. 14–15.
[31] G. Succi, W. Pedrycz, M. Marchesi, and L. Williams, "Preliminary analysis of the effects of pair programming on job satisfaction," in *Proceedings of the 3rd International Conference on Extreme Programming (XP)*, May 2002, pp. 212–215.
[32] P. Musílek, W. Pedrycz, N. Sun, and G. Succi, "On the Sensitivity of COCOMO II Software Cost Estimation Model," in *Proceedings of the 8th International Symposium on Software Metrics*, ser. METRICS '02. IEEE Computer Society, June 2002, pp. 13–20.
[33] M. Scotto, A. Sillitti, G. Succi, and T. Vernazza, "A Relational Approach to Software Metrics," in *Proceedings of the 2004 ACM Symposium on Applied Computing*, ser. SAC '04. ACM, 2004, pp. 1536–1540.
[34] J. W. Paulson, G. Succi, and A. Eberlein, "An empirical study of open-source and closed-source software products," *IEEE Transactions on Software Engineering*, vol. 30, no. 4, pp. 246–256, 2004.
[35] A. Sillitti, A. Janes, G. Succi, and T. Vernazza, "Measures for mobile users: an architecture," *Journal of Systems Architecture*, vol. 50, no. 7, pp. 393–405, 2004.
[36] M. Ronchetti, G. Succi, W. Pedrycz, and B. Russo, "Early estimation of software size in object-oriented environments a case study in a cmm level 3 software firm," *Information Sciences*, vol. 176, no. 5, pp. 475–489, 2006.
[37] M. Scotto, A. Sillitti, G. Succi, and T. Vernazza, "A non-invasive approach to product metrics collection," *Journal of Systems Architecture*, vol. 52, no. 11, pp. 668–675, 2006.
[38] R. Moser, W. Pedrycz, and G. Succi, "A Comparative Analysis of the Efficiency of Change Metrics and Static Code Attributes for Defect Prediction," in *Proceedings of the 30th International Conference on Software Engineering*, ser. ICSE 2008. ACM, 2008, pp. 181–190.
[39] ——, "Analysis of the reliability of a subset of change metrics for defect prediction," in *Proceedings of the Second ACM-IEEE International Symposium on Empirical Software Engineering and Measurement*, ser. ESEM '08. ACM, 2008, pp. 309–311.
[40] B. Rossi, B. Russo, and G. Succi, "Modelling Failures Occurrences of Open Source Software with Reliability Growth," in *Open Source Software: New Horizons - Proceedings of the 6th International IFIP WG 2.13 Conference on Open Source Systems, OSS 2010*. Notre Dame, IN, USA: Springer, Heidelberg, May 2010, pp. 268–280.
[41] W. Pedrycz, B. Russo, and G. Succi, "A model of job satisfaction for collaborative development processes," *Journal of Systems and Software*, vol. 84, no. 5, pp. 739–752, 2011.
[42] A. Sillitti, G. Succi, and J. Vlasenko, "Understanding the Impact of Pair Programming on Developers Attention: A Case Study on a Large Industrial Experimentation," in *Proceedings of the 34th International Conference on Software Engineering*, ser. ICSE '12. Piscataway, NJ, USA: IEEE Press, June 2012, pp. 1094–1101.
[43] E. Di Bella, A. Sillitti, and G. Succi, "A multivariate classification of open source developers," *Information Sciences*, vol. 221, pp. 72–83, 2013.
[44] L. Corral, A. B. Georgiev, A. Sillitti, and G. Succi, "Can execution time describe accurately the energy consumption of mobile apps? An experiment in Android," in *Proceedings of the 3rd International Workshop on Green and Sustainable Software*. ACM, 2014, pp. 31–37.
[45] I. D. Coman, P. N. Robillard, A. Sillitti, and G. Succi, "Cooperation, collaboration and pair-programming: Field studies on backup behavior," *Journal of Systems and Software*, vol. 91, pp. 124–134, 2014.
[46] L. Corral, A. Sillitti, and G. Succi, "Software Assurance Practices for Mobile Applications," *Computing*, vol. 97, no. 10, pp. 1001–1022, Oct. 2015.
[47] G. L. Kovács, S. Drozdik, P. Zuliani, and G. Succi, "Open Source Software for the Public Administration," in *Proceedings of the 6th International Workshop on Computer Science and Information Technologies*, October 2004.
[48] E. Petrinja, A. Sillitti, and G. Succi, "Comparing OpenBRR, QSOS, and OMM assessment models," in *Open Source Software: New Horizons - Proceedings of the 6th International IFIP WG 2.13 Conference on Open Source Systems, OSS 2010*. Notre Dame, IN, USA: Springer, Heidelberg, May 2010, pp. 224–238.
[49] B. Fitzgerald, J. P. Kesan, B. Russo, M. Shaikh, and G. Succi, *Adopting open source software: A practical guide*. Cambridge, MA: The MIT Press, 2011.
[50] B. Rossi, B. Russo, and G. Succi, "Adoption of free/libre open source software in public organizations: factors of impact," *Information Technology & People*, vol. 25, no. 2, pp. 156–187, 2012.
[51] Martin Fawler, "ShuHaRi," 2014. [Online]. Available: https://martinfowler.com/bliki/ShuHaRi.html
[52] M. S. Rahim, A. Chowdhury, D. Nandi, M. Rahman, and S. Hakim, "Scrumfall: A hybrid software process model," *International Journal of Information Technology and Computer Science*, vol. 10, pp. 41–48, 12 2018.
[53] L. Corral, A. Sillitti, G. Succi, A. Garibbo, and P. Ramella, "Evolution of Mobile Software Development from Platform-Specific to Web-Based Multiplatform Paradigm," in *Proceedings of the 10th SIGPLAN Symposium on New Ideas, New Paradigms, and Reflections on Programming and Software*, ser. Onward! 2011. New York, NY, USA: ACM, 2011, pp. 181–183.
[54] L. Corral, A. B. Georgiev, A. Sillitti, and G. Succi, "A method for characterizing energy consumption in Android smartphones," in *Green and Sustainable Software (GREENS 2013), 2nd International Workshop on*. IEEE, May 2013, pp. 38–45.